# Head phantoms for electroencephalography and transcranial electric stimulation: a skull material study

Alexander Hunold[1]*, Daniel Strohmeier[1], Patrique Fiedler[1], and Jens Haueisen[1,2]

[1] Institute of Biomedical Engineering and Informatics, Faculty of Computer Science and Automation, Technische Universität Ilmenau, Ilmenau, 98693, Germany

[2] Biomagnetic Center, Department of Neurology, Jena University Hospital, Jena, 07743, Germany

*Corresponding author:*

Alexander Hunold

Technical University Ilmenau

Institute for Biomedical Engineering and Informatics

POB 100565

D-98684 Ilmenau

Germany

Phone: +49 3677 69-1348

Fax: +49 3677 69-1311

Email: alexander.hunold@tu-ilmenau.de

**Abstract**

Physical head phantoms allow assessing source reconstruction procedures in electroencephalography and electrical stimulation profiles during transcranial electric stimulation. Volume conduction in the head is strongly influenced by the skull representing the main conductivity barrier. Realistic modeling of its characteristics is thus important for phantom development. In the present study, we proposed plastic clay as a material for modeling the skull in phantoms. We analyzed five clay types varying in granularity and fractions of fireclay, each with firing temperatures from 550 °C to 950 °C. We investigated the conductivity of standardized clay samples when immersed in a 0.9 % sodium chloride solution with time-resolved four-point impedance measurements. To test the reusability of the clay model, these measurements were repeated after cleaning the samples by rinsing in deionized water for 5 h. We found time-dependent impedance changes for approximately 5 min after immersion in the solution. Thereafter, the conductivities stabilized between 0.0716 S/m and 0.0224 S/m depending on clay type and firing temperatures. The reproducibility of the measurement results proved the effectiveness of the rinsing procedure. Clay provides formability, is permeable for ions, can be adjusted in conductivity value and is thus suitable for the skull modeling in phantoms.

**Abbreviations**

| | |
|---|---|
| EEG | Electroencephalography |
| NaCl | Sodium Chloride |
| TES | Transcranial Electric Stimulation |

**Introduction**

Volume conductor modeling of the human head can be used to calculate the distribution of electromagnetic fields caused by intracranial or extracranial generators, which is important for various applications in neuroscience. Electroencephalography (EEG) measures differences in the electric potential on the scalp generated by brain activity. Volume conductor modeling allows to solve the EEG forward problem, i.e., computing the EEG signals induced by neural activity in the brain. EEG source reconstruction methods use this information for solving the computationally complex, ill-posed EEG inverse problem to noninvasively estimate the location, orientation, and strength of the bioelectric sources in the brain [1]. These methods are routinely used in neuroscience to gain a better understanding of the brain functioning as well as in clinical applications, such as presurgical evaluation of epileptic activity [2]. Transcranial electrical stimulation (TES) is a noninvasive technique for brain stimulation. It is known to alter neural activity by means of cell membrane polarization. The effect of TES depends on the strength and orientation of the electric field in the target brain region [3]. Predictions of the electric field distribution in the brain are of great interest for obtaining dosage estimates and optimizing electrode positions and stimulation parameters for TES [4].

Verification and validation of the computational methods used for computing field distributions in the volume conductor as well as the estimation methods building on these forward models are essential steps in the development and application of novel EEG or TES approaches. For both tasks, physical phantom measurements provide advantages over in-vivo acquired data and computational simulations. Unlike in-vivo measurements, in phantoms physiological uncertainties do not exist and the ground truth in terms of source position, strength, orientation, and extent is known. In contrast to simulations, phantoms take into account real world influences, such as environmental noise or 3D positioning errors. Phantom investigations can be helpful in quality management, e.g., for the analysis of variability in EEG inverse procedures for different measurement sites and investigators. Here, physical phantoms allow for a comparison of the obtained source parameters for each site and investigator, including the complete measurement and analysis procedure, with the known ground truth in terms of source parameters. Another use case where phantoms are essential is the testing of detection limits for new equipment and experiments such as in neuronal current imaging [5].

Volume conduction in the head and the resulting distribution of electric potentials depends on the conductivity profile of the head. The compact skull, which is the head tissue with the lowest conductivity, is of particular relevance. Its conductivity is considerably lower than the conductivities of the surrounding tissues. Conductivity ratios for skull-to-brain conductivity range from 1/120 [6] to 1/5 [7] in the literature. Consequently, the practical realization of the skull compartment poses the most important and difficult task in head phantom construction. Realistic modeling of the electrolyte conductivity of the skull poses the challenge of realizing ion conduction in a complex geometry and simultaneously introducing a relevant structural conductivity barrier to ions.

In previous studies, head phantoms based on post-mortem human skulls were applied for the assessment of EEG source reconstruction procedures [8, 9]. These phantoms suffer from artificial conductivity adjustments to compensate denaturation.

Head phantoms based on compartments of saline solution were proposed for verifying simulations of TES [10, 11]. Kim et al. [10] generated a three compartment model based on saline doped agar solution comprising skin, skull and brain compartments. In this geometry, the skull compartment was modeled to unrealistically follow the folded surface of the brain. Further, surface ring electrodes were placed directly on the brain compartment, replicating brain stimulation. The resulting potential differences were sampled with depth electrodes inserted across all compartments. Similarly, surface copper electrodes mounted to the inside of a plastic skull model stimulating an inner compartment of saline solution was realized by Jung et al. [11].

In all previously listed models, compartments of different conductivities are realized by using saline solutions with varying ion concentration. Bringing compartments with different ion concentration in contact introduces diffusion interfaces. Diffusion of ions is a time dependent process and leads to instabilities of the initial conductivity configuration in a multi-compartment setup. Alternatively, homogeneous electrolyte concentration across compartments can be applied in combination with structural conductivity barriers provided by porous materials. This independence from electrolyte concentration gradients leads to temporal stability, especially in comparison to post-mortem skull [8, 9] and agar-based phantoms [10-12]. However, introducing a porous material as structural conductivity barrier into a saline solution leads to infiltration of the saline solution into the porous material. When the material is withdrawn from the saline solution, the aqueous component will evaporate and ions deposit in the porous material. This agglomeration of ions in the skull compartment can limit the reproducibility of the characteristics of the structural conductivity barrier provided by the porous material.

Materials applicable for modeling the skull in realistic EEG or TES head phantoms should allow replicating the geometry of the human skull and introduce a stable structural conductivity barrier.

In the present study, we analyzed clay as an electrolyte conductive material for modeling the skull in physical head phantoms for EEG and TES applications. In a multilayer head phantom, clay can provide a structural conductivity barrier allowing homogeneous electrolyte concentration across compartment layers, once the clay is infiltrated with saline solution comprising the electrolyte concentration of surrounding compartments. Sodium chloride (NaCl) solution supplies the charge carriers. Clay allows for arbitrary plastic designs due to its ductile deformability and provides reproducibility in its production process and therefore in its material properties, which can be exactly assessed. After firing, clay is dimensionally stable and allows cleaning procedures to remove ion agglomerations. Clay configurations vary with respect to their fraction of fire clay, their shrinkage and hygroscopicity. Further, the firing temperature influences the ceramics configuration and therefore microscopic structure. In this study, we seek to select optimal clay material and processing parameters for the skull compartment.

**Material and Methods**

*Clay types*

A federal ceramics consultant, Mr. Stefan Hasenöhrl (Technical Expert by the Chamber of Crafts, Erfurt, Germany), guided us in the selection of clay types with respect to varying hygroscopicity. The clay types further differ in their fraction of fire clay, granularity and shrinkage during the drying process. Fire clay is preheated clay that is ground or screened to its final particle size. As clay compound, the fraction of fire clay and its granularity influences the clay texture and its drying and shrinkage performance. Table 1 provides a categorization of the tested clay types shown in Figure 1A according to properties specified in technical data sheets.

**Table 1** Clay types ordered by identifier and categorized by material consistency, fire clay compound, shrinkage and hygroscopicity with property values according to technical data sheets.

| Type | Consistency | Fire Clay | | Shrinkage in % | | Hygroscopicity in % |
|---|---|---|---|---|---|---|
| | | Granularity in mm | Fraction in % | Drying | Firing | |
| 2sg | Stone ware compound | 0–5 | 60 | 6.9 | 2.5 (1000°C) | 10.3 (1000°C) |
| 264 | Modeling compound | 0–0.5 | 25 | n/a | 3 (1070°C) | 9 (1070°C) |
| 33 | Casting compound | n/a | n/a | 4.5 | 0.2 (1000°C) | 19.8 (1000°C) |
| 435 | Powdered clay | n/a | n/a | 2.6 | 0.1 (1000°C) | 14 (1000°C) |
| 1100 | Powdered clay | n/a | n/a | 5 | 3 (1070°C) | 12 (1070°C) |

The clay type 2sg (Arno Witgert Inh. Dipl.-Ing. (FH) Michael Liebig e.K, Herschbach, Germany) is a red-colored plastic body. The Creaton® types 264 and 1100 are throwing bodies and Creaton® types 33 and 435 are casting bodies (Creaton®, Goerg & Schneider GmbH u. Co. KG, Siershahn, Germany). The shrinkage fractions in Table 1 refer to the volumetric loss of the clay during either drying at room temperature (drying) or firing in a kiln (firing) at the respective firing temperature stated in brackets. The hygroscopicity values in Table 1 refer to the volume of water as a fraction of the clay volume, which can be absorbed by the clay after firing at the respective firing temperature given in brackets. For clay types containing no fire clay, specification of the granularity and fraction of the fire clay are not applicable (n/a). For each clay type, we generated nine different clay samples using firing temperatures from 550 °C to 950 °C with increments of 50 °C. Each clay sample was a disk with 74 mm in diameter and 6 mm in thickness. A representative example of each clay type is shown in Fig. 1A.

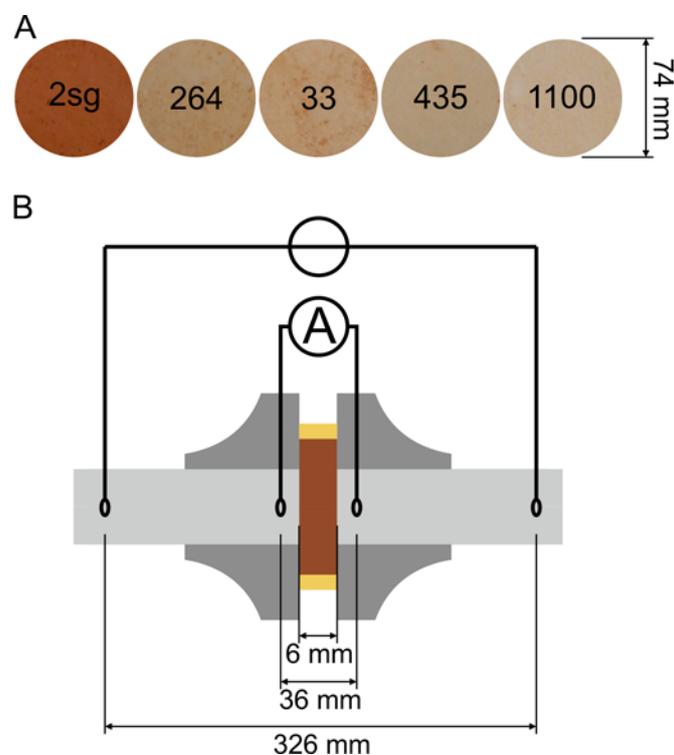

**Figure 1** Material samples and measurement scheme. A: Photos of five clay samples with labeled clay type. B: Impedance measurement scheme of the cell with a centered clay sample (red) surrounded by a silicon seal ring (yellow) placed between two composites of POM flanges (dark gray) and plastic tubes filled with NaCl solution (light gray) and two pairs of electrodes; the outer pair for voltage application driven by the voltage source and the inner pair for current measurement by the ampere meter, both being integrated into the Gamry Reference 600 impedance analyzer.

*Measurement setup*

We characterized the electric properties of the clay samples by means of time-resolved four-point impedance measurements. A Gamry Reference 600 (Gamry Instruments, Warminster, PA, USA) measured the impedance of the cell depicted in Figure 1B. In this context, cell denotes the combination of clay sample clamped between two NaCl solution compartments.

Each compartment held 750 ml of solution with a concentration of 0.9 % NaCl in deionized water. A silicone seal ring surrounded the clay sample when clamped between polyoxymethylene (POM) flanges. The flanges held DN 50 high-temperature resistant plastic effluent tubes. The electrodes were positioned in the center of the tube with distances of 15 mm and 160 mm to the clay sample respectively in both compartments. In the applied potentiostatic measurement mode, the outer pair of electrodes impressed the potential difference across the cell and the inner pair of electrodes measured the resulting current flow. The reliability of the measurement setup was verified in a leakage test and by verifying conductivity measurements of NaCl solutions performed with the digital conductivity meter HQ14D (Hach Lange GmbH, Duesseldorf, Germany).

*Impedance measurements*

Impedance measurements were performed at a constant frequency of 20 Hz over 25 min, where 61 single measurements were performed. Impedance spectroscopy was performed from 0.01 Hz to 10 kHz. Repeated impedance spectroscopy of 220 spectra in the range from 0.1 Hz to 10 kHz, with each spectrum lasting for 5 min, was performed to evaluate frequency dependency over time. Measurements were started immediately after immersion of the cleaned dry clay sample into the NaCl solution.

Impedances were measured across the whole cell comprising the NaCl solution and the clay sample. The difference between the impedance of a cell measurement and the impedance of a pure NaCl solution measurement as a reference provided the net impedance for each sample. From the net impedance $Z$ of the sample, we calculated the sample conductivity $\sigma$ according to:

$$\sigma = \frac{l}{Z*A} \qquad (1)$$

with the clay sample thickness $l$ and the surface area $A$.

*Cleaning procedure*

Considering the reusability of head phantoms, we evaluated the reproducibility of the electrochemical characteristics. After immersing a porous clay sample in NaCl solution, NaCl deposits in the sample's structure during the storing period before the next measurement. Consequently, a cleaning procedure for the clay samples was implemented. Clay samples with deposited NaCl were immersed into deionized water and rinsed for 5 h under constant stirring. During this time period, the rinsing water was refreshed every 30 min, yielding 10 cleaning cycles.

**Results**

*Impedance measurements*

During the first 4 min of the impedance measurements, the measured cell impedance dropped about 2 decades. Measurements of 0.9 % NaCl solution revealed an impedance of 47 Ohm. The difference between the cell impedance and the impedance of the NaCl solution leads to the impedance of the clay sample. An example of impedance traces for cell impedance, the impedance of the NaCl solution and the net impedance for sample 33 at 950 °C is depicted in Figure 2A. Figure 2B presents the calculated conductivity trace for sample 33 at 950 °C. In order to evaluate the conductivity of a certain sample, we considered the last value of the 25 min interval.

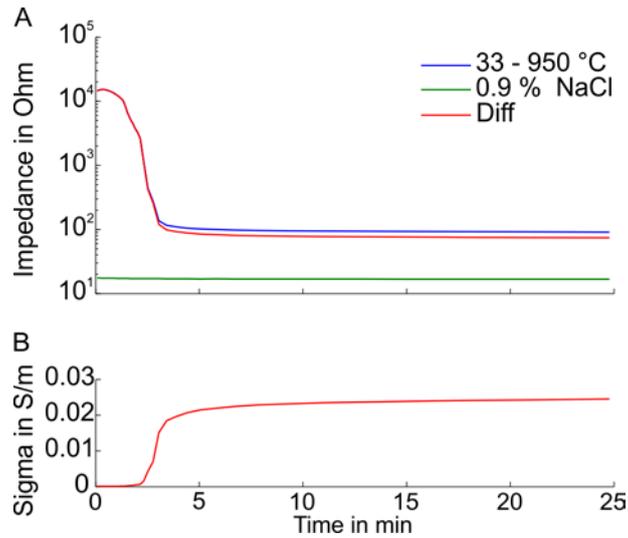

**Figure 2** Impedance measurement data and calculated conductivity over time. A: Impedance data over 25 min for 0.9 % NaCl solution (green) and the clay sample of type 33 with firing temperature 950 °C in 0.9 % NaCl solution (blue). The red line visualizes the net impedance of the clay sample as the difference between the blue and green traces. B: Calculated conductivity from net impedance of the aforementioned clay sample.

The tested clay samples revealed conductivity values between 0.191 S/m and 0.024 S/m with an overall average of 0.048 S/m (Figure 3). The samples of the clay types without fire clay marked the boundaries of highest and lowest electrolyte conductivity. The clay type with high shrinkage and low hygroscopicity, type 1100, presented the highest conductivity of all samples with mean value and standard deviation across firing temperatures of 0.153±0.016 S/m. The powdered clay 1100 with firing temperature 900 °C provided the highest conductivity overall of 0.191 S/m. The clay types with low shrinkage and high hygroscopicity, types 33 and 435, presented lowest conductivities with mean values and standard deviations across the firing temperatures of 0.045±0.016 S/m and 0.056±0.023 S/m. The casting compound 33 with firing temperature 950 °C provided the lowest conductivity of 0.024 S/m. The conductivity values of clay with fractions of fire clay, types 2sg and 264, ranged with 0.079±0.008 S/m and 0.102±0.019 S/m as mean values and standard deviations across the firing temperatures in-between the conductivity values of the other clay types.

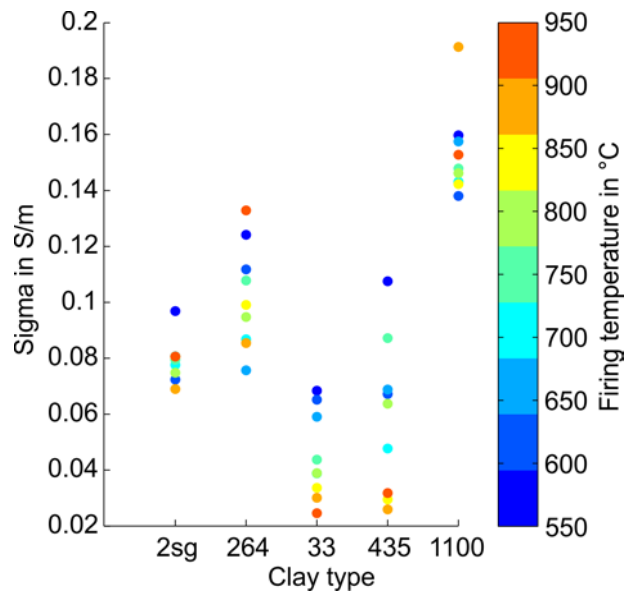

**Figure 3** Conductivities of all clay samples according to clay type and color-coded with respect to samples' firing temperature. Conductivity values represent the calculation results from the stabilized impedance value at the end of a measurement interval of 25 min.

Figure 4 presents the conductivity over time for all samples of the clay types 33 and 1100 as these types presented clearly distinct clusters of conductivity values. For the samples of clay type 33, the conductivity consistently decreased with increasing firing temperature. In samples of other clay types, including type 1100, the firing temperature and the conductivity were not directly related. Further, samples of clay type 33 provided more stable conductivity values compared to samples of clay type 1100. During the last 10 min of the measurement interval conductivity values increased 0.0011±0.0004 S/m in samples of clay type 33 and 0.0051±0.0027 S/m in samples of clay type 1100.

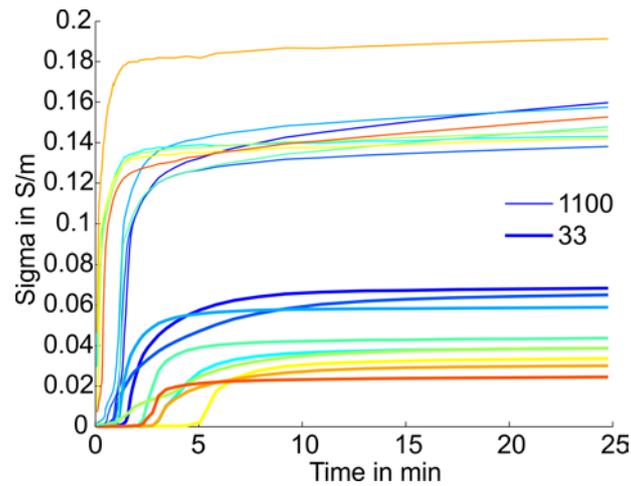

**Figure 4** Conductivity over time for clay types 33 (thick lines) and 1100 (thin lines). Colors code the firing temperature according to the color bar in Figure 3.

Repeated impedance spectroscopy measurements of a sample of clay type 264 with firing temperature 550 °C comprising 220 spectra in the range from 0.1 Hz to 10 kHz with each spectrum lasting for 5 min revealed no frequency dependent impedance changes.

*Cleaning procedure*

Repeated application of our cleaning procedure between measurement repetitions resulted in stable sample conductivity across repeated impedance spectrograms of the sample of clay type 33 with firing temperature 950 °C (Figure 5). The quantitative analysis of the calculated conductivity in the range from 10 Hz to 10 kHz provided a mean conductivity of 0.024 ± 0.0006 S/m (mean ± std) over the four measurements.

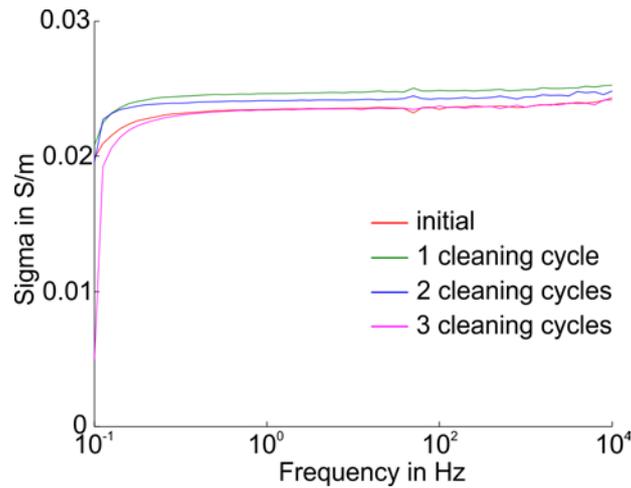

**Figure 5** Conductivity spectra calculated from repeated impedance measurements of the clay sample of type 33 with firing temperature 950 °C. Red: initial impedance measurement; Green, blue, magenta: impedance measurement after performing the cleaning procedure and drying the sample, once, twice and three times.

**Discussion**

In this study, we investigated fired clay samples with respect to their potential use as skull material in a physical head phantom for EEG and TES. We investigated five clay types with desired plastic formability, different hygroscopicity, shrinkage and fire clay compound. From each clay type, nine samples with different firing temperatures were analyzed. Overall, the conductivity for the clay samples in NaCl solution was within the range of approximately 0.02 S/m to 0.2 S/m after the initial settling time of approximately 5 min. Repeated impedance spectroscopy measurements provided evidence for the frequency independence of the conductivity of the clay samples in the range of 0.1 Hz to 10 kHz.

Conductivity values ranging from 0.00275 S/m [6] to 0.066 S/m [7] were used in previous investigations for the human skull. An often indicated skull conductivity was 0.022 S/m [13]. The conductivity values of the tested clay samples in the present study included a value of 0.024 S/m for the sample of clay type 33 with a firing temperature of 950 °C, deviating only 9 % from this literature value. Further, the clay type 33 covered a conductivity range from 0.068 S/m to 0.024 S/m monotonically decreasing with increasing firing temperature. Consequently, this clay type, especially the sample with firing temperature 950 °C, represents a potential skull modeling material with respect to its properties as structural conductivity barrier.

A common coarse approximation of the head as volume conductor comprises three homogeneous conductivity compartments, representing the soft tissues inside the skull, the skull, and the scalp [14]. A value of 0.33 S/m [15] is widely used to describe the conductivity of all tissues inside the skull compartment, which is also used as a reference for the conductivities of the other compartments. The skull bone represents the main conductivity barrier compared to the surrounding soft tissue layers, specified by skull-to-brain conductivity ratios from 1/120 [6] to 1/5 [7] in the literature. Consequently, modeling and source localization studies [14, 16] revealed a dominant influence of the skull layer. Thus, clay types in this study cover ratios from approx. 1/14 to approx. 1/2, in particular depending on the firing temperature samples of clay type 33 covered ratios from approx. 1/14 to approx. 1/5.

Phantoms designed for combined near-infrared spectroscopy (NIRS) and EEG as introduced by Cooper et al. [17, 18] and Barbour et al. [19], are based on polymer resins doped by either gold-coated copper wires [17] or saline [19] to adapt to the physiological conductivity. However, both phantoms only model two compartments, not explicitly addressing the skull as the main conductivity barrier.

A phantom for electrical impedance tomography introduced a skull modeling approach facilitating agar hydrogel [12]. In this study, an agar solution based on distilled water was used to model the skull compartment. To achieve an appropriate conductivity value, the skull was modeled with a thickness of 25 mm, distorting the geometric proportion. Preventing diffusion of saline across different agar compartments, Sperandio et al. placed a volume conductive film between the phantom components. However, this study only considered frequencies above 1 kHz, which are beyond the main temporal dynamic range of bioelectric generators and TES.

Physical phantoms with ion conduction on physiological scale demonstrated feasibility to synthetic EEG and TES analysis [9, 11]. Another head phantom for EEG verification was based on carbon doped silicon and urethane resin [20]. In this phantom, the conductivity mechanism relied on electron conduction in contrast to the ion conduction in human tissue. When applying an EEG measurement setup, an electrolyte conductive gel was introduced to mediate the interface between the measurement electrode and skin compartment. Consequently, two electrode–ion interfaces built up and introduced diffusion dependent processes.

However, in layered phantoms, comprising connected compartments with different ion concentrations, the diffusion of ions might limit the phantom stability [21-23]. Multi-compartment phantoms implementing a constant ion concentration throughout all compartments and realizing different conductivities by means of structural barriers would overcome this limitation. With respect to the temporal performance of the clay samples, our results demonstrate a stable conductivity after a transient interval of a few minutes (c.f. Figures 2 and 4). The transient interval demonstrates the decrease in cell resistance due to ion infiltration into the clay sample. In contrast, the resistance of the homogeneous reference cell containing only NaCl solution was constant throughout the measurement duration. Consequently, the conductivity trace (Figure 2B) derived from the difference between the complete cell and the reference cell (blue curve in Figure 2) also showed the transient interval, with increasing conductivity due to the infiltration of the clay sample by charge carriers. Differences in the structural properties across all tested clay types led to variations in the duration of the transient interval and in the stability of the conductivity value as demonstrated in the comparison of clay types 33 and 1100 in Figure 4. The conductivity of clay type 33 with a firing temperature of 950 °C changed for only 2.5 % during the last 10 min in the measurement interval. Consequently, this sample with the desired structural conductivity demonstrated a higher stability than the average of this, in general stable, clay type.

The clay materials used in this study were commercially available clay compounds or powders. A ceramics master and authorized expert in the field of pottery produced the clay samples according to a standard procedure. Using standardized clay types processed by an experienced ceramics master, we can ensure reproducibility of the clay samples. For the production of complex shaped skull models, newly developed direct ink writing methods with ceramic compounds might be feasible [24].

Clay represents a natural material which underwent a manual production procedure for generation of the samples investigated in this study. Consequently, sample properties, i.e. the area $A$ and the thickness $l$, used in equation (1) had tolerances influencing the calculated conductivity values. Across the presented samples, the standard deviation of the conductivity due to geometric variation was approx. 10 %. Taking this variation into account, our results still represent conductivity values relevant for skull modeling with respect to literature values [14-16].

Clay can be successfully applied for modeling the skull compartment and represents its structural conductivity barrier characteristic. Soaking the clay material with NaCl ensures the ionic conductivity. As part of a multi-compartment phantom, the skull compartment interacts with the adjacent compartments, e.g. scalp and intracranial soft tissue. In order to guarantee reproducible phantom measurement conditions, the mechanically stable skull compartment needs to be reconfigured to the initial state. We implemented and validated a cleaning procedure for rinsing deposits, i.e., NaCl deposits, from the used clay samples. Our results demonstrate a reproducible conductivity configuration in repeatedly measured and cleaned clay samples (c.f.

Figure 5). Consequently, the skull compartment of a head phantom modeled with clay material could be reused.

**Conclusion**

We investigated the applicability of fired clay for modeling the skull in physical head phantoms for EEG and TES. Clay is a well-known, formable material available at low cost, which is inherently stable and permeable for ions after firing. Our measurements showed that fired clay provides a stable conductivity barrier allowing for physiologically plausible skull conductivity values depending on the clay type and firing temperature. Furthermore, clay samples provided adjustable electrolyte conductivity with respect to both initialization and reproducibility. Among the tested set of clay configurations, we found the sample of type 33 with a firing temperature of 950 °C, generating a conductivity of 0.024 S/m, most suitable for modeling the skull compartment. The proposed cleaning procedure ensured reproducible conductivity characteristics and thus reusability of the skull model.

**Acknowledgements**

We would like to thank the Technical Expert by the Chamber of Crafts, Erfurt, Germany, Mr. Stefan Hasenöhrl, who supported us in the selection of clay types and manufactured the clay samples. This project has received funding from the European Union's Horizon 2020 research and innovation programme under grant agreement No 686865 (project BREAKBEN), from the Thuringian Ministry of Economic Affairs, Science and Digital Society under FGR 0085 (project BASIs), and from the German Research Foundation under grand number HA 2899/21-1.

# References

[1] S. Baillet, J. C. Mosher, and R. M. Leahy. Electromagnetic brain mapping. IEEE Signal Processing Magazine 2001; 18: 14-30.

[2] A. Hunold, J. Haueisen, B. Ahtam, C. Doshi, C. Harini, S. Camposano et al. Localization of the Epileptogenic Foci in Tuberous Sclerosis Complex: A Pediatric Case Report. Frontiers in Human Neuroscience 2014; 8.

[3] M. A. Nitsche, and W. Paulus. Excitability changes induced in the human motor cortex by weak transcranial direct current stimulation. The Journal of physiology 2000; 527: 633-639.

[4] S. Wagner, F. Lucka, J. Vorwerk, C. S. Herrmann, G. Nolte, M. Burger et al. Using reciprocity for relating the simulation of transcranial current stimulation to the EEG forward problem. Neuroimage 2016; 140: 163-173.

[5] N. Hofner, H. H. Albrecht, A. M. Cassara, G. Curio, S. Hartwig, J. Haueisen et al. Are brain currents detectable by means of low-field NMR? A phantom study. Magn Reson Imaging 2011; 29: 1365-73.

[6] S. Homma, T. Musha, Y. Nakajima, Y. Okamoto, S. Blom, R. Flink et al. Conductivity ratios of the scalp-skull-brain head model in estimating equivalent dipole sources in human brain. Neuroscience Research 1995; 22: 51-55.

[7] K. Wendel, N. G. Narra, M. Hannula, P. Kauppinen, and J. Malmivuo. The Influence of CSF on EEG Sensitivity Distributions of Multilayered Head Models. IEEE TRANSACTIONS ON BIOMEDICAL ENGINEERING 2008; 55: 1454-1456.

[8] R. M. Leahy, J. C. Mosher, M. E. Spencer, M. X. Huang, and J. D. Lewine. A study of dipole localization accuracy for MEG and EEG using a human skull phantom. Electroencephalography and clinical Neurophysiology 1998; 107: 159-73.

[9] S. Baillet, J. J. Riera, G. Marin, J. F. Mangin, J. Aubert, and L. Garnero. Evaluation of inverse methods and head models for EEG source localization using a human skull phantom. Physics in Medicine and Biology 2001; 46: 77-96.

[10] D. Kim, J. Jeong, S. Jeong, S. Kim, S. C. Jun, and E. Chung. Validation of computational studies for electrical brain stimulation with phantom head experiments. Brain stimulation 2015: 1-35.

[11] Y. J. Jung, J. H. Kim, D. Kim, and C. H. Im. An image-guided transcranial direct current stimulation system: a pilot phantom study. Physiol Meas 2013; 34: 937-50.

[12] M. Sperandio, M. Guermandi, and R. Guerrieri. A four-shell diffusion phantom of the head for electrical impedance tomography. IEEE TRANSACTIONS ON BIOMEDICAL ENGINEERING 2012; 59: 383-9.

[13] T. F. Oostendorp, J. Delbeke, and D. F. Stegeman. The conductivity of the human skull: results of in vivo and in vitro measurements. IEEE TRANSACTIONS ON BIOMEDICAL ENGINEERING 2000; 47: 1487-92.

[14] M. Stenroos, A. Hunold, and J. Haueisen. Comparison of three-shell and simplified volume conductor models in magnetoencephalography. Neuroimage 2014; 94: 337-48.

[15] L. A. Geddes, and L. E. Baker. The specific resistance of biological material--a compendium of data for the biomedical engineer and physiologist. Med Biol Eng 1967; 5: 271-93.

[16] Ü. Aydin, J. Vorwerk, M. Dümpelmann, P. Küpper, H. Kugel, M. Heers et al. Combined EEG/MEG Can Outperform Single Modality EEG or MEG Source Reconstruction in Presurgical Epilepsy Diagnosis. PLoS ONE 2015; 10.

[17] R. J. Cooper, R. Eames, J. Brunker, L. C. Enfield, A. P. Gibson, and J. C. Hebden. A tissue equivalent phantom for simultaneous near-infrared optical tomography and EEG. Biomedical Optics Express 2010; 1: 425-430.

[18]   R. J. Cooper, D. Bhatt, N. L. Everdell, and C. H. Jeremy. A tissue-like optically turbid and electrically conducting phantom for simultaneous EEG and near-infrared imaging. Physics in Medicine and Biology 2009; 54: N403.

[19]   R. L. Barbour, H. L. Graber, Y. Xu, Y. Pei, C. H. Schmitz, D. S. Pfeil et al. A programmable laboratory testbed in support of evaluation of functional brain activation and connectivity. IEEE Transactions on Neural Systems and Rehabilitation Engineering 2012; 20: 170-83.

[20]   T. J. Collier, D. B. Kynor, J. Bieszczad, W. E. Audette, E. J. Kobylarz, and S. G. Diamond. Creation of a human head phantom for testing of electroencephalography equipment and techniques. IEEE TRANSACTIONS ON BIOMEDICAL ENGINEERING 2012; 59: 2628-2634.

[21]   R. J. Sadleir, F. Neralwala, T. Te, and A. Tucker. A Controllably Anisotropic Conductivity or Diffusion Phantom Constructed from Isotropic Layers. Annals of biomedical engineering 2009; 37: 2522-2531.

[22]   F. Wetterling, M. Liehr, P. Schimpf, H. Liu, and J. Haueisen. The localization of focal heart activity via body surface potential measurements: tests in a heterogeneous torso phantom. Physics in Medicine and Biology 2009; 54: 5395.

[23]   U. Tenner, J. Haueisen, H. Nowak, U. Leder, and H. Brauer. Source localization in an inhomogeneous physical thorax phantom. Physics in Medicine and Biology 1999; 44: 1969.

[24]   J. A. Lewis, J. E. Smay, J. Stuecker, and J. Cesarano. Direct ink writing of three-dimensional ceramic structures. Journal of the American Ceramic Society 2006; 89: 3599-3609.